\documentclass[prl,twocolumn,superscriptaddress,showpacs]{revtex4}
\usepackage{amssymb,amsmath}

\def\As{{\mathcal A}}
\def\Bs{{\mathcal B}}
\def\cA{\mathcal{A}}
\def\cB{\mathcal{B}}
\def\cH{\mathcal{H}}

\def\ie{\textit{i.e. }}

\begin{document}
\title{Quantum Statistics and Locality}
\author{Detlev Buchholz}  
\affiliation{Institut f\"ur Theoretische Physik, 
Universit\"at G\"ottingen,  37077 G\"ottingen, Germany} 
\author{Stephen J.\ Summers}
\affiliation{Department of Mathematics, 
University of Florida, Gainesville FL 32611, USA}
\begin{abstract} \noindent
It is shown that two observers have mutually commuting observables if 
they are able to prepare in each subsector of their common state 
space some state exhibiting no mutual correlations.
This result establishes a heretofore missing link between 
statistical and locality (commensurability) properties of the observables of 
spacelike separated observers in relativistic quantum physics, envisaged
four decades ago by Haag and Kastler. It is based on a discussion
of coincidence experiments and suggests a physically
meaningful quantitative measure
of possible violations of Einstein causality.
\end{abstract}
\pacs{02.30.Tb, 03.65.Ud, 11.10.Cd}
\maketitle 
\noindent Within the setting of relativistic quantum 
physics, the principle of Einstein causality is commonly expressed by the
assumption that the observables of spacelike separated observers
commute. This postulate, often called the condition of
locality \cite{Haag}, has become one of the basic ingredients
in both the construction and the analysis of relativistic 
theories \cite{We}. Yet, in spite of its central role in the 
theoretical framework, the question of whether locality can be 
deduced from physically meaningful properties of the 
physical states has been open for more than four decades.

     Indeed, in 1964 Haag and Kastler \cite{HaKa} proposed an approach to
this question using a condition of statistical independence, now commonly 
termed $C^*$-independence \cite{Su90}. Roughly speaking, it says that
two spacelike separated 
observers should find no obstructions when preparing arbitrary 
states in their respective spacetime localization regions, 
despite the actions of the fellow observer. Thus,  
such pairs of partial states should arise by restricting 
some global state to the observable algebras of the respective
observers. Though this property is certainly of physical relevance, 
it turned out not to ensure the desired locality \cite{Ek,FaTa}. 
Related notions of statistical independence
were proposed in \cite{Kr,Li}, and some formulations
of independence motivated by quantum logic were examined later in 
\cite{Re95a,Re95b}. Although of physical interest, 
all these concepts were determined to be logically independent of locality.
Certain exceptions in this respect are the approaches in 
\cite{MuEi,Na} which, however, do not go to the
heart of the matter. For recent reviews of this subject and
further references to the literature, cf. \cite{Su90,Re98}. 

     It is the purpose of this note to propose a condition 
on the set of common states of two spacelike separated observers 
which is based upon the coincidence experiments such  
observers can sensibly perform. As we shall see, our condition 
implies that the observable algebras of these observers satisfy 
locality and a strong version of Haag--Kastler independence. 
Moreover, our approach suggests a quantitative measure
of possible violations of Einstein causality.

Proceeding to the mathematical formulation, 
let $\As$, respectively $\Bs$, 
represent the algebra of observables of the 
two respective observers and let $\cA \bigvee \cB$ be
the algebra generated by all of these observables. 
Without essential loss of generality we may assume that these
algebras are weakly closed subalgebras of the algebra of all bounded
operators on some Hilbert space $\cH$, each of which containing the unit
operator $1$. The states (expectation functionals) $\omega$ on these
algebras are, for any operator $C$, given by the familiar formula
\begin{equation*}
\omega(C) = \text{Tr} \, \rho \, C \, , 
\end{equation*}
where $\rho$ is some density matrix on $\cH$.  Thus, technically
speaking, we adopt here the standard formulation of quantum physics in
terms of von Neumann algebras and their respective normal states.

     Since we want to discuss the nature of the correlations between
observables $A \in \cA$ and $B \in \cB$ in the given states, we have
to define a product describing the observables of corresponding
coincidence experiments. As is well known, the ordinary product of $A$
and $B$ is not suitable for this purpose, since it is not
even hermitian if the operators $A$ and $B$ do not commute. 
The symmetrized (Jordan)
product is better behaved in this regard.  But it is also not
acceptable as a coincidence observable, since, for example, the 
Jordan product of positive operators need not be positive.

     In order to understand what is physically required, let us 
discuss the case of ``yes--no'' experiments described by orthogonal
projections $E \in \cA$ and $F \in \cB$.  Thus, in a single
measurement, each observer finds either one of the two possible values
$1, 0$.  If the observers decide to make a corresponding 
coincidence experiment, they have to design a ``yes--no'' experiment
such that whenever the apparatus gives a reading of $1$
then any subsequent measurement of $E$, respectively $F$, must 
yield the value $1$ with certainty. Moreover, they must
strive to maximize the acceptance rate of the device.
The corresponding observable is  
the orthogonal projection $E \wedge F$, \ie the largest
projection contained in both $E$ and $F$. We recall that this
projection can be obtained as the (strong) limit, $\lim_{n \rightarrow
\infty} (EF)^n = \lim_{n \rightarrow \infty} (FE)^n = E \wedge F$.
It coincides with the ordinary product if the
projections commute, and it is zero if the properties fixed
by these projections are completely incompatible.
In a similar manner one can define a coincidence product
$A \wedge B$ of arbitrary observables $A$, $B$, but we make no
use of this notion here.

     Now we turn to the discussion
of the nature of the correlations between observables in $\cA$ and
$\cB$ in the underlying states. We begin with a formal definition of
uncorrelated states.\\[1mm]
\noindent 
\textbf{Definition:} A state $\omega$ on $\cA \bigvee \cB$ as described
above is said to be $\cA$--$\cB$--uncorrelated if $\omega(E\wedge F) =
\omega(E) \, \omega(F)$, for every pair of projections $E \in \cA$ and
$F \in \cB$. \\[1mm]
\indent It follows from the preceding discussion that this definition
can, in principle, be checked by the
described observers.  Let us discuss the consequences of the
assumption that such an uncorrelated state exists. Here we denote,
as is customary, the projections $1-E$, $1-F$ onto the complements of
given projections $E$, $F$ by $E^\perp$ and $ F^\perp$, respectively.
Making use of the linearity of states and the assumption that $\omega$
is $\cA$--$\cB$--uncorrelated, we obtain for all $E,F$ as indicated
\begin{equation*} 
\begin{split}
& \omega(E \wedge F + E \wedge F^\perp + E^\perp \wedge F
+  E^\perp \wedge F^\perp ) \\ 
& = \omega(E + E^\perp) \, \omega(F + F^\perp) = 1 \, ,
\end{split}
\end{equation*} 
where the last equality follows from the fact
that $\omega$ is normalized. The orthogonal projection
\begin{equation*}
T(E,F) \doteq E \wedge F + E \wedge F^\perp + E^\perp \wedge F
+  E^\perp \wedge F^\perp
\end{equation*}
appearing in this equation will be called the total coincidence of $E$
and $F$. It is invariant if one replaces $E,F$ by their respective
complements $E^\perp$ and $F^\perp$. From the preceding result, 
$\omega(T(E,F)) = 1$, it
follows that there is some non--zero vector $\Phi \in \cH$, lying in
the range of the density matrix $\rho$ of $\omega$, such that
\begin{equation*}
T(E,F) \, \Phi = \Phi \, , 
\end{equation*}
for every pair of projections $E \in \cA$, $F \in \cB$. Now
\begin{equation*}
E F \, T(E,F) = E \wedge F = F E \, T(E,F) \, ,
\end{equation*}
and consequently $EF \, \Phi = FE \, \Phi$. Since this equality holds
for every pair of projections, it extends, by the spectral theorem, to
arbitrary pairs of observables in $\cA$ and $\cB$. Taking also into
account the fact that every operator in $\cA$ and $\cB$ can be
decomposed into a (complex) linear combination of observables in the
respective algebra, we thus find that
\begin{equation*}
X Y \, \Phi = Y X \, \Phi, \qquad X \in \cA, \, Y \in \cB \, .
\end{equation*}
Next, making use of this equation several times, we obtain 
for $X, X_1 \in \cA$ and $Y, Y_1 \in \cB$
\begin{equation*}
\begin{split}
& X Y \, X_1 Y_1 \Phi = X (Y Y_1) X_1 \Phi = (X X_1) (Y Y_1) \Phi \\
& = (Y Y_1) (X X_1) \Phi = Y (X X_1) Y_1 \Phi = Y X \, X_1 Y_1 \Phi \, .
\end{split}
\end{equation*}
Similarly, 
for arbitrary operators $X, X_1, \dots, X_n \in \cA$
and $Y, Y_1, \dots, Y_n \in \cB$, we obtain by induction in $n$
\begin{equation*}
\begin{split}
& X Y \, X_1 Y_1 \cdots X_n Y_n \Phi \\
& = X Y \, X_1 Y_1 \cdots X_{n-1} (Y_{n-1} Y_n) X_n \Phi \\
& = X Y \, X_1 Y_1 \cdots (X_{n-1} X_n) (Y_{n-1} Y_n) \Phi \\
& = Y X \, X_1 Y_1 \cdots (X_{n-1} X_n) (Y_{n-1} Y_n) \Phi \\
& = Y X \, X_1 Y_1 \cdots X_{n-1} (Y_{n-1} Y_n) X_n \Phi \\
& = Y X \, X_1 Y_1 \cdots X_n Y_n \Phi \, .
\end{split}
\end{equation*}
So we conclude that the elements of the algebras $\cA$ and $\cB$
commute with each other on the subspace of vectors obtained by
applying all elements of the algebra $\cA \bigvee \cB$ to the vector
$\Phi$.

     It is an immediate consequence of this result that for any vector
$\Psi$ in this subspace and any pair of projections as above
one has
\begin{equation*}
E \wedge F \, \Psi = \lim_{n \rightarrow \infty} (EF)^n \, \Psi
= E F \, \Psi \, .
\end{equation*}
This implies in particular that all of these vectors satisfy the same
basic relation as the vector $\Phi$ with which we began, \ie
\begin{equation*}
T(E,F) \, \Psi = \Psi \, ,
\end{equation*}
for every pair of projections $E \in \cA$, $F \in \cB$.

     Now let $\cH_p \subset \cH$ be the subspace generated by all
vectors $\Psi$ satisfying the latter condition and let $P$ be
the orthogonal projection onto this space.  Thus $P$ is the largest
projection which is contained in all projections $T(E,F)$ with $E$,
$F$ as before.  In particular, $P$ is an element of the algebra $\cA
\bigvee \cB$. On the other hand, it follows from the preceding
discussion that $P$ commutes with all elements of $\cA \bigvee \cB$, 
since $\cH_p$ is stable under the action of this algebra. So $P$ is an
element of the center of $\cA \bigvee \cB$.

     If $P \neq 1$, the subsector $\cH_p^\perp = (1 - P) \cH$ is
non--trivial, and there is no density matrix in this space giving rise
to an $\cA$--$\cB$--uncorrelated state.  Hence, assuming that the two
observers are in a position to prepare uncorrelated states in every
subsector of their common state space, we are led to the conclusion
that $P = 1$. So we have arrived at our main result, relating
statistical and locality properties of the observables. \\[1mm]
\noindent \textbf{Theorem:} Consider two observers with 
observable algebras $\cA$ and $\cB$, respectively. If these 
observers can prepare in each subsector of their common
state space some $\cA$--$\cB$--uncorrelated state, 
the elements of their algebras commute with each other.  \\[1mm]
     As a matter of fact, more can be said about the structure of the
algebra $\cA \bigvee \cB$, when the hypothesis of this theorem holds.
In order to simplify this discussion, let us assume that $\cA \bigvee
\cB$ has trivial center. Then it follows from the theorem and a result
in \cite{Tak} that the algebra $\cA \bigvee \cB$ is isomorphic to the
von Neumann tensor product $\cA \otimes \cB$. In fact, 
this isomorphism is implemented by a unitary operator
\cite{DaLo}. This implies in
particular that the algebras $\cA$ and $\cB$ are statistically
independent in the sense that, given any pair $\omega_a$, $\omega_b$
of normal states on $\cA$ and $\cB$, respectively, there is some
normal product state $\omega$ extending this pair:
\begin{equation*}
\omega(A B) = \omega_a(A) \, \omega_b(B), \quad
A \in \cA, B \in \cB \, .
\end{equation*}
As the algebras $\cA$ and $\cB$ commute with each other, any product
state is clearly also $\cA$--$\cB$--uncorrelated, so such states exist
in abundance. 

The general case, where $\cA \bigvee \cB$ has a non--trivial
center, can always be reduced to the one just
discussed by restricting the algebra $\cA \bigvee \cB$ to subsectors
where all central elements act as multiples of the identity (central
decomposition). Therefore, under the conditions 
described in the theorem, not only does locality follow but
also a strong form of statistical independence. We regard 
the combination of these features as the appropriate expression of 
the principle of Einstein causality.

     It should be noted that these results are not in conflict with the
existence of $\cA$--$\cB$--entangled states.  In fact, all that
matters in this context is the quantum nature of the individual
algebras $\cA$ and $\cB$.  If these algebras contain copies of the
Pauli matrices (which is always the case if they are of infinite type)
there exist states on $\cA \bigvee \cB$ which maximally violate Bell's
inequalities in correlation experiments \cite{SuWe}.  Hence, the
so--called quantum--nonlocality of entangled states and the locality
of observable algebras are two perfectly compatible aspects of quantum
physics, even though this twofold usage of the term ``local'' 
has led to confusion.

     Since violations of Einstein causality are theoretically discussed and
may even materialize experimentally should quantum effects of gravity
become accessible, we conclude with some remarks on a quantitative
measure of such deviations. Under the favorable circumstances 
described above, two spacelike separated observers are able to avoid
correlations between their subsystems. A natural measure for 
these $\cA$--$\cB$--correlations in a given state $\omega$
is the quantity
\begin{equation*}
C_\omega(\cA,\cB) \doteq 
\sup_{E,F} |\omega(E \wedge F) - \omega(E) \omega(F)| \, ,
\end{equation*}
where the supremum extends over all pairs of projections
in the respective algebras. It is zero in all
$\cA$--$\cB$--uncorrelated states $\omega$.  Thus one may take 
\begin{equation*}
C(\cA,\cB) \doteq \inf_\omega C_\omega(\cA,\cB) \, 
\end{equation*}
as a measure of the violation of causality in a given 
sector, the infimum being taken over all corresponding
states. The value of this quantity  
ranges between $0$ and $1$. One extreme corresponds
to complete independence of the given algebras
and the other to maximal incommensurability of the underlying 
observables. The latter case is realized, for example,  
in quantum mechanics by the algebras generated by the 
position and momentum operator, respectively. 
It is straightforward to provide examples of noncommuting
algebras for which $0 < C(\cA,\cB) < 1$.

A violation of Einstein causality for spacelike separated
observers would thus manifest itself in a value of $C(\cA,\cB)$
different from zero. For macroscopic distances of 
the observers, this value should become minute, however.
One may therefore expect that essential portions of the 
physical interpretation of local theories, such as 
the description of particles, their statistics, collision theory 
\textit{etc}. should still apply in an asymptotic sense in the presence of 
such acausal effects. Further analysis of these aspects of 
the notions and results established in the present letter
seems warranted.

\vspace*{2mm}

\begin{acknowledgments}
\noindent 
We are grateful to M.~R\'edei for pointing out to us
reference \cite{MuEi} and to R.~Werner for illuminating remarks.
DB also wishes to thank the Department of Mathematics and 
the Institute for Fundamental Theory of the University
of Florida for hospitality and financial support.
\end{acknowledgments}

\end{document}